\documentclass[sigconf, review]{acmart}

\renewcommand\footnotetextcopyrightpermission[1]{} 

\AtBeginDocument{%
  \providecommand\BibTeX{{%
    \normalfont B\kern-0.5em{\scshape i\kern-0.25em b}\kern-0.8em\TeX}}}

\def\ie{\textit{i.e.}}
\def\etal{\textit{et al.} }
\def\etc{\textit{etc.} }

\def\eg{\textit{e.g.}}

\def\wrt{\textit{w.r.t. }}
\def\aka{\textit{a.k.a. }}

\usepackage{multirow}

\usepackage{amssymb}
\usepackage{amsmath}
\usepackage{appendix} 

\begin{document}

\title{Know in \emph{AdVance}: Linear-Complexity Forecasting of Ad Campaign Performance with Evolving User Interest}





\author{XiaoYu Wang\footnotemark[1], YongHui Guo\footnotemark[4], Hui Sheng\footnotemark[4], Peili Lv\footnotemark[4], Chi Zhou\footnotemark[4], Wei Huang\footnotemark[4], ShiQin Ta\footnotemark[4], Dongbo Huang\footnotemark[4], XiuJin Yang\footnotemark[4], Lan Xu\footnotemark[4], Hao Zhou\footnotemark[2], and Yusheng Ji\footnotemark[1]} 
\affiliation{ 
  \institution{\footnotemark[1]National Institute of Informatics, Tokyo, Japan \ \footnotemark[4]Tencent \  \footnotemark[2]University of Science and Technology of China}
   \country{China}
}
\email{wangxiaoyu1001@gmail.com} 
\email{{brookguo, hughsheng, paleylv, fredchizhou, johnwhuang, secondta, andrewhuang, xiujinyang, lanxu}@tencent.com}
\email{kitewind@ustc.edu.cn}
\email{kei@nii.ac.jp}

\begin{abstract}

Real-time Bidding (RTB) advertisers wish to \textit{know in advance} the expected cost and yield of ad campaigns to avoid trial-and-error expenses. However, Campaign Performance Forecasting (CPF), a sequence modeling task involving tens of thousands of ad auctions, poses challenges of evolving user interest, auction representation, and long context, making coarse-grained and static-modeling methods sub-optimal. We propose \textit{AdVance}, a time-aware framework that integrates local auction-level and global campaign-level modeling. User preference and fatigue are disentangled using a time-positioned sequence of clicked items and a concise vector of all displayed items. Cross-attention, conditioned on the fatigue vector, captures the dynamics of user interest toward each candidate ad. Bidders compete with each other, presenting a complete graph similar to the self-attention mechanism. Hence, we employ a Transformer Encoder to compress each auction into embedding by solving auxiliary tasks. These sequential embeddings are then summarized by a conditional state space model (SSM) to comprehend long-range dependencies while maintaining global linear complexity. Considering the irregular time intervals between auctions, we make SSM's parameters dependent on the current auction embedding and the time interval. We further condition SSM's global predictions on the accumulation of local results. Extensive evaluations and ablation studies demonstrate its superiority over state-of-the-art methods. AdVance has been deployed on the Tencent Advertising platform, and A/B tests show a remarkable 4.5\% uplift in Average Revenue per User (ARPU).

\end{abstract}


\maketitle

\pagestyle{plain} 

\section{Introduction}
\label{sec:intro}

Online display advertising, especially the dominant Real-time Bidding (RTB) paradigm, has evolved into a \$300 billion market \cite{dentsu2022globalspend} and becomes the primary revenue source for tech giants such as Google, Meta, Alibaba, Tencent, etc. Its success lies in a \textit{win-win} situation: platforms monetize user visits into the opportunities of displaying ads (\aka \textbf{user impression}), while advertisers purchase such impressions to reach potential customers and promote marketing. RTB allows advertisers to pre-define certain criteria for launching ad campaigns. Criteria specify bid prices, target audience (\eg, females aged 20-35 living in Shanghai), and optimization objectives (\eg, pursuing more exposure, clicks, or conversions). Then, a long series of auctions competing for the user impressions that satisfy such criteria constitutes the ad campaign. 
 
RTB features \textit{non-guaranteed} delivery (NGD) modes, \ie, both the cost and yield of a campaign remain uncertain before its fulfillment. Consequently, it is critical for advertisers to \textbf{\textit{know in advance}} the expected performance, rendering the Campaign Performance Forecasting (CPF) problem. This foresight brings two-fold benefits: 1) Advertisers use a few tentative predictions to balance a wider audience and higher conversion rates, thus improving Return on Investment (ROI). 2) Platforms can stimulate advertisers to invest additional budget for more yield, thus promoting revenue. 

CPF problem induces a \textit{sequence-to-sequence} task, with the input of an auction series satisfying the campaign criteria, and the output of the corresponding cost and yield so far. Significant academic and industry attention has been attracted: Kalish \etal from Bidtellect \cite{kalish2016method} constructed a multi-variate time series of similar campaigns to predict new campaigns. Wu \etal from Tencent \cite{wu2021efficient} estimated a scaling factor of the total future impression volume. These \textit{coarse-grained} methods fail to harness the information of each auction. In contrast, Wang \etal from Yahoo \cite{yahoo-wang2009search} aggregated qualified auctions from bid logs, and Jiang \etal from Meta \cite{facebook-jiang2015predicting} further considered reaching distinct users, albeit lacking \textit{a global viewpoint} from campaign-level modeling. Nath \etal from Microsoft \cite{microsoft-nath2013ad} combined dynamic linear models with Bayes net for winning price estimation. Cui \etal \cite{yahoo-cui2013campaign} used probabilistic methods of a mixed log-normal distribution, while Ren \etal \cite{ren2019deep} replaced it with recurrent neural networks (RNN) to approximate a discrete winning distribution. However, neglecting \textit{evolving user interest} results in a substantial gap between predictions and online results.

To fill this gap, we propose \emph{AdVance}, a time-aware framework that integrates local and global modeling. Designing such a framework faces the following challenges: 

\begin{enumerate}

    \item \textbf{Evolving user interest: }During the exposure to a series of ads, a user clicks the preferred ads and accumulates fatigue toward the similar ones, causing future click and conversion rates to decline. This accounts for the \emph{diminishing marginal utility} issues where the yield is not proportional to the cost increment. Neglecting this phenomenon renders over-estimated campaign performance. 
    
    
    \item \textbf{Auction representation: }Each auction involves user features, contextual information, and a dynamic number of candidate ads competing with each other. The platform's filtering rules further complicate the auction process. Effectively compressing and extracting useful information from such a \textit{multi-source}, \textit{variable-length} input remains a significant challenge.
       
    \item \textbf{Long context: }Accurate predictions require summarizing a sequence of tens of thousands of auctions with irregular time intervals. Traditional linear architectures like RNNs and CNNs struggle to model long-range dependency, while the self-attention mechanism suffers quadratic complexity, making it impractical for CPF tasks. 
\end{enumerate}

AdVance converts each auction and corresponding user interest into a single embedding. It summarizes the embedding sequence with a conditional State Space Model (SSM) to achieve linear complexity. Specifically, we use a time-positioned click sequence and a fatigue vector compressing all displayed ads to reflect interest dynamics. A Transformer encoder conducts self- and cross-attention on candidate ads and user-related features and predicts the auction-level cost and yield. This fully utilizes the supervision signals from historical records and forges an informative representation. SSMs feature linear structures like RNNs and CNNs while achieving comparative long-range modeling ability as self-attention. We propose its conditional variant. We condition its parameters on the current auction and time interval to handle the irregular input sequence, and we condition the campaign-level prediction on the accumulated auction-level outputs.

In summary, our contributions are as follows:

\begin{itemize}
    \item We focus on the challenging task of forecasting ad campaign performance with evolving user interest, which benefits both advertisers and platforms by providing valuable insights and stimulating ad budgets. 

    \item We propose \emph{AdVance}, a time-aware framework that combines auction- and campaign-level modeling. AdVance leverages the attention mechanism to vectorize each auction locally and summarizes the whole sequence with a conditional SSM, achieving global linear complexity. 

    \item We conduct evaluations and ablation studies using large-scale industrial datasets, demonstrating the superiority of AdVance over state-of-the-art methods. AdVance has been deployed on the Tencent advertising platform, and we uploaded the PyTorch implementation. \footnote{\url{https://github.com/anonymousauthor113/CPF}}
\end{itemize}

\section{related work}
\label{sec:section2}

\subsection{Campaign Performance Forecasting}

Accurate modeling of campaigns grants advertisers insights into their investment and return, thus attracting significant research interests.  Based on the granularity, existing works can be categorized into campaign-level and auction-level methods. Kalish \etal \cite{kalish2016method} estimated campaign performance by aggregating statistics from similar historical campaigns. Wu \etal \cite{wu2021efficient} focused on calculating scaling coefficients to adjust predicted volumes of future impressions to earned ones. Despite having low complexity, they discard fine-grained information within each auction, leading to a non-negligible accuracy gap. 

Auction-level methods, in contrast, lift the complexity for higher forecasting accuracy, as the benefits for publishers and advertisers are significant. Wang \etal \cite{yahoo-wang2009search} estimated a quality score for each (ad, user)-pair using regression modeling and used it as a threshold to select qualified impressions. Cui \etal \cite{yahoo-cui2013campaign} enhanced this work by incorporating probabilistic methods and assuming a mixture of log-normal distribution. Jiang \etal  \cite{facebook-jiang2015predicting} calculated corresponding threshold bid prices for winning historical auctions and counted the number of exposed users. Chen \etal \cite{alibaba-chen2022unified} further augmented it with multi-task learning and campaign information. 

Following the spirit of estimating threshold prices to win, another line of research known as \textit{bid landscape} or \textit{market price modeling} has gained traction and can be applied to campaign modeling tasks. As a representative,  Ren \etal \cite{ren2019deep} removed assumptions on the distribution forms and utilized a recurrent neural network to flexibly model the conditional winning probability for each bid price. Yang \etal \cite{yang2021multi} further incorporated multi-task learning to jointly model click-through rate and market price, thereby providing multiple results in a single return to enhance the robustness and online inference efficiency. 

The main drawback of these methods is the neglect of user interest evolvement \textit{in the future campaign environment} and directly using historical click/conversion probability. When a particular ad wins more auctions, the user preference evolves, and fatigue accumulates toward repetitive similar ads. Neglecting such evolvement and assuming static user interest causes an overestimate of campaign performance and budget waste. 

\subsection{User Interest Modeling}

User interest modeling mainly focuses on the probability of certain explicit behaviors, such as clicking or conversion, by modeling the feature interaction between users and ads. Early machine-learning and deep-learning methods, including logistic regression \cite{richardson2007predicting}, gradient boosting decision trees (GBDT) \cite{he2014practical}, collaborative filtering \cite{sarwar2001item}, Wide\&Deep \cite{widedeep-cheng2016wide}, DeepFM \cite{guo2017deepfm}, DCN \cite{deepcross-wang2017deep}, and PNN \cite{PNN-qu2016product}, adopt a \textit{static} viewpoint and overlook the dynamics of user preference. To address the limitation, DIN \cite{din-zhou2018deep} first incorporated the sequence of the user's historic clicked items and utilized an attention mechanism to build an enriched user feature. Subsequently, a series of works such as DIEN \cite{dien-zhou2019deep}, DSIN \cite{dsin-feng2019deep}, SIM \cite{SIM-pi2020search}, UBR4CTR \cite{UBR4CTR-qin2020user}, SMR\cite{SIM-pi2020search} emerged to model user interest evolution using recurrent neural network (RNN) \cite{hochreiter1997long} or Transformers \cite{vaswani2017attention}. However, the aforementioned methods discard the abundant displayed but \textit{non-clicked} ads, which account for user fatigue towards repeated similar ads. In contrast, AdVance considers all displayed ads to comprehensively understand interest evolution.

\section{Preliminary}
\label{sec:section3}

\subsection{Attention Mechanism}

Attention mechanism \cite{vaswani2017attention} excels at modeling long-range dependencies. It allows different parts (\aka tokens) of the input sequence to interact, regardless of their distance and position. This is achieved by representing the input \textit{Queries} as the weighted sum of \textit{Values}. The weights depend on the similarity between queries and \textit{Keys}, measured by the dot-product: 
\begin{align}
    \label{eq:attention}
    & \textrm{Attn}(\textbf{Q}, \textbf{K}, \textbf{V}) = \textrm{softmax}(\frac{\textbf{Q}\textbf{K}^\intercal}{\sqrt{d_K}}) \textbf{V}, 
\end{align}
where $d_K$ is the dimension of each key vector. 

For the self-attention, $\textbf{Q} = \textbf{XW}_Q$, $\textbf{K} = \textbf{XW}_K$, and $\textbf{V} = \textbf{XW}_V$ are the projections of the \textbf{\textit{same}} sequence $\textbf{X}$, thereby focusing on information exchange and aggregation within single sequence. 

In contrast, the cross-attention involves two \textbf{\textit{different}} sequences $\textbf{X}$ and $\textbf{Y}$, where $\textbf{Q} = \textbf{XW}_Q$, $\textbf{K} = \textbf{YW}_K$, and $\textbf{V} = \textbf{YW}_V$. This allows $\textbf{X}$ to ``borrow" information from $\textbf{Y}$, thus useful in multi-modality learning such as vision-language models \cite{alayrac2022flamingo, rombach2022stablediffusion}. 

The attention mechanism's main drawback is its quadratic complexity, as each new token has to attend to \textbf{\textit{all}} previous tokens. This incurs heavy burdens for handling numerous ad auctions. 

\subsection{State Space Model}

As a promising competitor to Transformers, the State Space Model (SSM) \cite{gu2022S4} shares the same virtue of linear recurrence of RNN while achieving comparative long-range modeling capacity in sequence analysis \cite{smith2023s5}, time series prediction \cite{zhang2023effectively}, and large language models \cite{fu2023H3, gu2023mamba}. It defines a continuous differential system and recurrently updates a hidden state $h(t)$: 
\begin{align}
    \frac{\mathrm{d}h}{\mathrm{d}t} = \textbf{A}h(t) + \textbf{B}x(t), \quad y(t) = \textbf{C}h(t), 
    \label{eq:ssm}
\end{align}
where $x(t) \in \mathcal{R}$ is the 1-D input signal, $y(t) \in \mathcal{R}$ is the output signal, $\textbf{A} \in \mathcal{R}^{N \times N}$ is the state transition matrix, $\textbf{B} \in \mathcal{R}^{N \times 1}$ is the input matrix, and $\textbf{C} \in \mathcal{R}^{1 \times N}$ is the output matrix. 

To adapt Eq. \eqref{eq:ssm} to sequence modeling tasks, we employ zero-order hold (ZOH), a technique for discretizing continuous equations, and we have the linear recurrence form: 
\begin{align}
    h_t = \overline{\textbf{A}}h_{t-1} + \overline{\textbf{B}}x_t, \quad y_t = \textbf{C}h_t, 
    \label{eq:discretization}
\end{align}
where $\overline{\textbf{A}} = \exp(\Delta \textbf{A})$, $\overline{\textbf{B}} = (\Delta \textbf{A})^{-1}(\exp(\Delta \textbf{A}) - \textbf{I}) \cdot \Delta \textbf{B}$, and $\Delta$ denotes the step size. Note that $h_t$, $x_t$, and $y_t$ are now discrete time series. 

When the input is a sequence of $D$-dimension vectors, we stack $D$ SSMs to model each vector dimension separately, resulting in a total $(ND)$-dimension hidden space. Like the Transformers, a multi-layer perceptron (MLP) is often added to process the concatenation of all $D$ SSMs' output (\aka channel mixing \cite{yu2022metaformer}). 

\subsection{Problem Formulation}
\label{problem-formulation}
Advertisers pre-define the criteria of ad campaigns to expose target users to their ads within a specific period. An auction is launched whenever a qualified user impression comes, and $\sim200$ selected candidates compete for it. Eventually, an irregular time series of auctions constitutes the campaign. 

We define \textbf{campaign performance} as the expected cost and yield of an ad campaign. Advertisers may pursue more ad exposure, clicks, or conversions, giving rise to CPM (Cost-per-Mille), CPC (Cost-per-Click), and CPA (Cost-per-Action) ad types. The \textbf{expected yield} of an auction is defined as: 
\begin{align}
    \label{eq:ecpm}
    \textrm{yield} = 
    \begin{cases}
        \textrm{win-rate} \times 1&\textrm{CPM} \\ 
        \textrm{win-rate} \times \textrm{pCTR} &\textrm{CPC} \\ 
        \textrm{win-rate} \times \textrm{pCVR} &\textrm{CPA}
    \end{cases}
\end{align}
Here, win-rate is the target ad's probability of winning the auction, pCTR (predicted click-through rate) is the probability of the user clicking the ad, and pCVR (predicted conversion rate) denotes the probability of the user's conversion like adding to cart or purchase. Then we define the \textbf{expected cost} of such an auction as $(\textrm{bid price} \times \textrm{expected yield})$. This is also known as the effective cost-per-mille (eCPM). 

Given advertiser-defined criteria and a long sequence of qualified auctions, our target problem is to predict the corresponding cost and yield of the campaign with evolving user interest and maintain acceptable algorithm complexity for practical needs.

\section{Method}
\label{sec:section4}

\begin{figure*}
    \centering
    \includegraphics[width=0.95\linewidth]{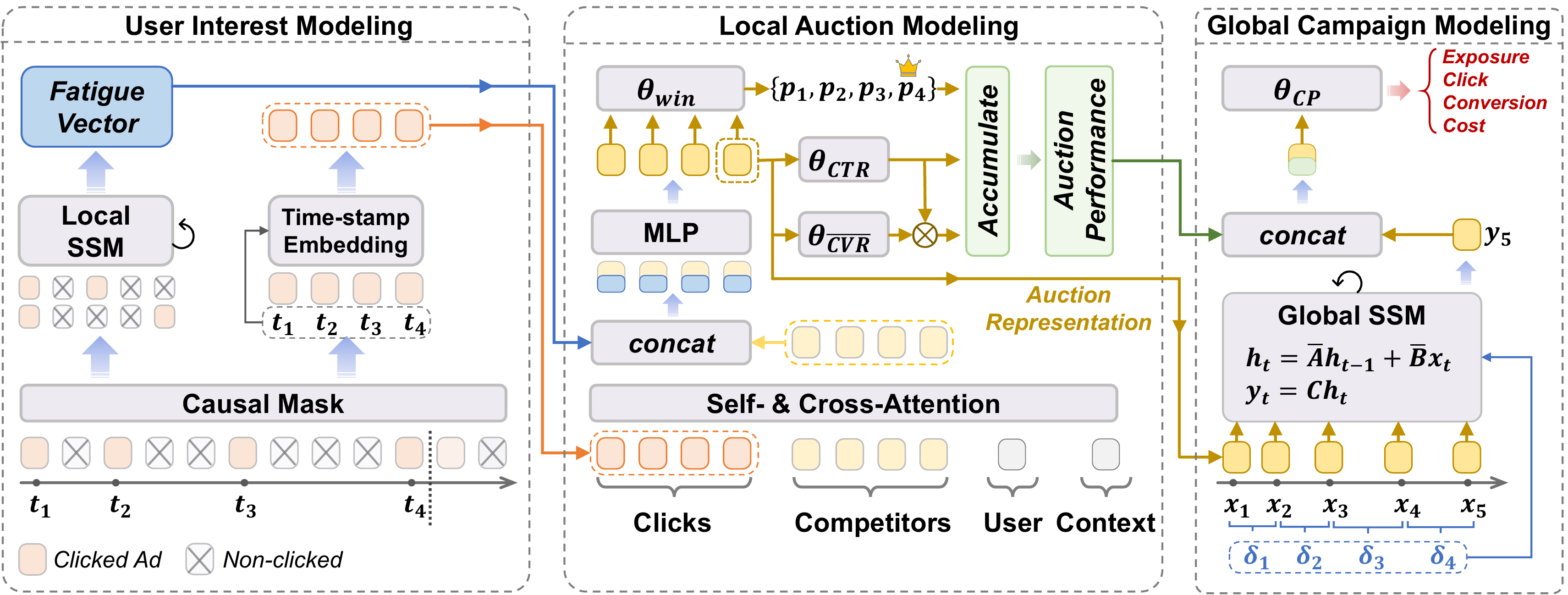}
    \caption{AdVance disentangles user interests as time-stamped click sequences representing user preference and fatigue vectors compressing all displayed items (Sec. \ref{user_interest}). The attention mechanism compresses auctions into dense embeddings, and AdVance accumulates auction-level performance (Sec. \ref{local_auction}). A global SSM recurrently summarizes all embeddings, and AdVance returns final results based on the summary and accumulated performance (Sec. \ref{global_SSM}). During training, a causal mask blocks out ``future" records after the current time-stamp (Sec. \ref{offline_training}). }
    \label{fig:AdVance}
\end{figure*}

AdVance operates on a sequence-to-sequence paradigm by mapping a series of auctions to a series of estimated campaign performances at the moment, as illustrated in Fig. \ref{fig:AdVance}. AdVance consists of three modules, \ie, user interest, local auction, and global campaign modeling. Click records with time-stamp embedding reflect user preference. The local SSM recurrently updates the fatigue vector based on the whole display history. An encoder conducts self- and cross-attention on candidate ads and user features to predict auction performance, thereby fully utilizing the log data and building an informative representation. The generated sequence of auction embedding has long lengths and irregular time intervals. The linear-complexity, global SSM with parameters dependent on inputs and intervals summarizes the sequence. The final prediction relies on the global SSM's output and the accumulated auction performance, thereby tightly integrating the fine-grained auction and holistic campaign knowledge.  

\subsection{User Interest Modeling}
\label{user_interest}
The sequence of a user's previous clicks and conversions offers a basis for estimating the preference towards similar ads. Many works \cite{din-zhou2018deep, dien-zhou2019deep, dsin-feng2019deep} incorporate it as part of user features, albeit with two drawbacks. 

\begin{itemize}
    \item \textbf{Irregular interval: }RNN- and Transformer-based methods treat user behaviors as an evenly distributed time series, while the interval lengths affect the \textbf{timeliness} of the relevance between historical and current behaviors (\eg, a purchase made a week ago is often more informative than one made two months ago). 

    \item \textbf{User fatigue: }Displayed yet non-clicked records account for fatigue accumulation and yield declines. Ignoring them causes overestimated preferences and displaying similar ads repeatedly. 
\end{itemize}

Recent behaviors deserve more emphasis for modeling user interest, as discovered by \cite{guo2023query, li2020time}. At the Tencent advertising platform, each display record has a time-stamp. Inspired by the absolute and relative positional embedding \cite{vaswani2017attention, shaw2018self}, we propose relative time-stamp positional embedding as: 
\begin{align}
    \textrm{Pos}(t) = [\cos(\omega_1 t), \sin(\omega_1 t), \dots, \cos(\omega_d t), \sin(\omega_d t)], 
    \label{eq: position_embedding}
\end{align}
where $t$ is the difference between the current time-stamp and a manually set origin (\eg, 2023.1.1 0:00 AM), $2d$ equals the dimension of input embedding, and $\omega_1, \dots, \omega_d$ are $d$ trainable parameters. We calculate $\textrm{Pos}(t)$ for each click record and add it to the click record's embedding. We use trigonometric functions due to their good properties for dot-product: 
\begin{align}
    \textrm{Pos}(t) \cdot \textrm{Pos}(t +\delta) = \cos(\omega_1 \delta) + \cos(\omega_2 \delta) + \dots + \cos (\omega_d \delta), 
\end{align}
where $\delta$ represents a time interval. Therefore, the time distance information is preserved for AdVance to pay attention to more relevant user behaviors. 

A seemingly feasible solution to handle non-clicked records is to include them as user features, just like what we do to the clicked ones. However, the number of non-clicked is usually 20 or more times larger than that of clicked records, making it impractical due to the quadratic complexity of the self-attention mechanism. 

We employ a local state space model (SSM) to compress the whole sequence of displayed ads into a fatigue vector in linear complexity, which serves as conditional information in the auction representation (Sec. \ref{local_auction}). It processes the display records one by one and recursively updates the fatigue vector. We make the SSM's parameters data-dependent and interval-dependant, granting the model the ability to \textit{selectively} memorize salient knowledge from the irregular input series. Sec. \ref{global_SSM} provides more details about the conditional SSM. 

\subsection{Local Auction Modeling}
\label{local_auction}

This module takes an input of click sequence, fatigue vector, a \textbf{\textit{varying}} number of candidate ads, user profile, and other contextual information to learn an informative representation for each auction. This demands the model architecture capable of 1) handling variable-length input, 2) modeling competitive relations between any two of the candidates, and 3) aggregating knowledge from multiple input sources into one vector. 

We employ an attention-based encoder that satisfies the aforementioned demands. The encoder conducts self-attention on the candidate ads, where the competitive relationship forms a complete graph. The encoder applies cross-attention between candidate ads and the rest of the input to extract knowledge from user profiles, interests, and context. This knowledge indicates the user's value to advertisers. Formally, we have: 
\begin{align}
    \textbf{X} &= \textbf{X} + \textrm{Pos}(t), \notag \\
    \textbf{X} &= \textrm{LN}(\textbf{X} + \textrm{Attn}(\textbf{XW}_Q, [\textbf{X}; \textbf{Y}]\textbf{W}_K, [\textbf{X}; \textbf{Y}]\textbf{W}_V)), \notag \\
    \textbf{X} &= \textrm{LN}(\textbf{X} + \textrm{MLP}(concat(\textbf{X}, \Vec{f}))), 
\end{align}
where $\textbf{X}$ denotes the candidate ad embeddings, and $\textbf{Y}$ denotes the embeddings of user click sequence, user profile, and contextual information. We calculate $\textrm{Pos}(t)$ using the time-stamp of the current auction. We use $[\textbf{X}; \textbf{Y}]$ to calculate the keys and values, thus integrating the self- and cross-attention in one pass. $\textrm{LN}$ denotes the layer norm function \cite{layernorm-ba2016layer}, and $\textrm{MLP}$ represents a multi-layer perceptron, often stacked fully-connected layers with ReLU activation as in \cite{vaswani2017attention}. We use the $concat(\cdot)$ operator to concatenate the fatigue vector $\Vec{f}$ to \textbf{each} ad embedding, as this factor greatly affects user clicks and conversions. 

Empirically, supervised learning is a more straightforward and sample-efficient paradigm for representation learning \cite{mahajan2018exploring}. The common practice is to first train a model on a labeled dataset, then remove the classifier (usually the last few layers of the model). Then, the rest of the model serves as a discriminative representation extractor. This inspires us to devise a \textbf{multi-task} architecture of predicting each auction's win-rate and expected yield, with the shared representation of ad embedding $\textbf{X}$. 

\textbf{Win-rate prediction: }Besides bid prices and user-ad matching, the Tencent platform manually sets filtering rules that can not be described as analytic functions. Inspired by PointerNet \cite{vinyals2015pointer} and AlphaStar \cite{vinyals2019grandmaster}, we treat the auction process as a black box and approximate it with a discrete distribution over all ads. We train a \textit{win-rate model} $f(\cdot; \theta_{win})$ to compress \textit{each} ad embedding $X_i$ into a scalar $w_i$ that describes its relative advantage over other ads. A Softmax layer then turns all scalars into a discrete distribution of the winning probability $p_i = \exp(w_i) / \sum_{j=1, 2, \dots} \exp(w_j)$ for each ad. The ground truth is recorded as a one-hot vector $[0 \dots 1 \dots 0]$, where $1$ indicates the winner. Hence, we use the categorical cross-entropy as our loss function. 

\textbf{Yield prediction: }We focus on estimating pCTR and pCVR (Sec. \ref{problem-formulation}). We model the task as a binary classification problem and use a Sigmoid function $f(x) = 1/(1+e^{-x})$ to output a probability. However, predicting pCVR faces a great challenge due to the sparser positive samples than the pCTR problem. Inspired by ESMM \cite{esmm-ma2018entire}, we introduce two sub-models $f(\cdot; \theta_{CTR})$ and $f(\cdot; \theta_{\overline{CVR}})$: the former predicts pCTR, and the latter predicts the conversion probability \textbf{conditioned} on that the ad has been clicked. Apparently, the ad conversion must come after the ad click. Thus, $f(\cdot; \theta_{CTR}) \times f(\cdot; \theta_{\overline{CVR}})$ equals pCVR, according to the chain rule. This design lowers the difficulty of learning pCVR by treating pCTR as an intermediate task and solving a conditional probability problem in a smaller space. Furthermore, it allows AdVance to output multiple yield metrics of exposure, click, and conversion volumes with Eq. \eqref{eq:ecpm}, regardless of the campaign objectives. 

Notably, win-rate and yield prediction are correlated tasks, as ads with high pCTR/pCVR also have a higher rate of winning the auction. This connection benefits their training mutually and helps to learn a more effective representation, as discovered by \cite{yang2021multi} and our experiments. \textbf{We select the target ad's embedding as the auction representation}, as it has aggregated information from all other tokens after the cross- and self-attention. 

\subsection{Global Campaign Modeling}
\label{global_SSM}

This module takes input from a time series of auction embedding and summarizes it into a \textbf{summary vector}. Then, AdVance forecasts the campaign performance based on such a vector. 

Self-attention models preserve all previous tokens as \textit{Key} and \textit{Value} matrices, and each new token has to traverse the sequence before it, resulting in a quadratic complexity. In contrast, State Space Models (SSMs) maintain a hidden state to \textit{compress} historical input. This allows SSMs to process new tokens recurrently and update the hidden state correspondingly, thus achieving a linear complexity. 

However, the vanilla SSM's parameters $(\Delta, \textbf{A}, \textbf{B}, \textbf{C})$ remain the same for all tokens \cite{gu2022S4}. A constant stepsize $\Delta$ is unsuitable for irregular auction intervals, and a static input matrix $\textbf{B}$ and output matrix $\textbf{C}$ can not selectively preserve or discard information based on the current token, causing a redundant hidden state.

Inspired by the gating mechanism \cite{hochreiter1997long, cho2014GRU, hua2022FLASH}, recent research suggests a data-dependent design that makes SSM's parameters as functions of input tokens \cite{smith2023s5, gu2023mamba}. Therefore, we define the \textbf{conditional SSM} as 
\begin{align}
    \textbf{B} &= \textbf{X}W_{\textbf{B}} + b_{\textbf{B}}, \notag \\
    \textbf{C} &= \textbf{X}W_{\textbf{C}} + b_{\textbf{C}}, \notag \\
    \Delta &= \tau_{\Delta} \big( concat(\textbf{X}, \delta_{\textbf{X}})W_{\Delta} + b_{\Delta} \big). 
\end{align}
Here, $\textbf{X} = [x_1, x_2,\dots] \in \mathcal{R}^{L \times D}$ denotes an $L$-length sequence of $D$-dimension auction embedding. $W_{\textbf{B}}, W_{\textbf{C}} \in \mathcal{R}^{D \times N}$ map input tokens into the input matrix and output matrix, respectively. $\delta_{\textbf{X}}$ denotes time intervals between the successive auctions. Its first entry is set to $0$. We concatenate  $\textbf{X}$ and $\delta_{\textbf{X}}$ along the dimension axis, thereby making AdVance aware of the \textbf{time irregularity}. $W_{\Delta} \in \mathcal{R}^{(D+1) \times D}$ maps input tokens and time intervals into the SSM's stepsizes, and $\tau_{\Delta}(x) = \log(1 + \exp(x))$ denotes the Softplus function, a smooth approximation to the ReLU function, making sure the stepsizes always positive. $b_{\textbf{B}}$, $b_{\textbf{C}}$, and $b_{\Delta}$ are all biases. 

Following the same setting as \cite{gu2023mamba, zhang2023effectively, ma2022mega}, we set the transition matrix $\textbf{A} \in \mathcal{R}^{N \times N}$ as diagonal to save computation. Note that the hidden state's dimension $N$ is often much smaller than $D$. To slim $W_{\Delta} \in \mathcal{R}^{(D+1) \times D}$, we replace it by the product of $W_{\Delta}^{(1)} \in \mathcal{R}^{(D+1) \times N}$ and $W_{\Delta}^{(2)} \in \mathcal{R}^{N \times D}$, reducing the $\mathcal{O}(D^2)$ to $\mathcal{O}(ND)$. 

Once we get $(\Delta, \textbf{A}, \textbf{B}, \textbf{C})$, we calculate the discretized version $(\overline{\textbf{A}}, \overline{\textbf{B}}, \textbf{C})$ using Eq. \eqref{eq:discretization} and train the conditional SSM. Multiple techniques can accelerate AdVance's training speed, such as FlashAttention \cite{dao2022flashattention} and parallel scan \cite{blelloch1990prefix}. 

We use the SSM's last output $y_L$ as the whole campaign's summary vector. We also accumulate each auction's expected cost, exposure, click, and conversion and concatenate them into a vector $P_{accu} \in \mathcal{R}^4$. Finally, AdVance predicts all metrics of  campaign-level performance in one pass as a 4D vector:  
\begin{align}
    [cost, exp, clk, cvr] = f \big(concat(y_L, P_{accu}); \theta_{CP} \big), 
\end{align}
where $\theta_{CP}$ is the model parameter. 

\subsection{Training and Inference}

\subsubsection{Offline Training}
\label{offline_training}
Each training sample corresponds to one logged ad campaign. It contains a time-stamped sequence of auctions in which this ad has participated. Each auction sample records all competitor ads, user-related features, contextual information, the auction winner, and the users' click/conversion behavior. To lower variance and stabilize model convergence, we split the input sequence into chunks of 100 auctions. The training label is a time series of the corresponding total cost and yields up to that moment and is also aggregated by chunks. 

The training follows a Seq2Seq paradigm \cite{sutskever2014sequence}: AdVance sequentially processes input auctions and predicts the current campaign performance whenever a chunk has been finished. The loss is calculated \wrt the labels, and AdVance updates its parameters using back-propagation. Note that the global SSM only takes gradients \wrt campaign performances, while the auction-level and user-interest models take gradients from both campaign performance and auxiliary tasks. This creates a mini-batch training for the win-rate and yield prediction models. To prevent label leakage from the user's display history, we devise a \textbf{causal mask} that ``covers" the records after the current time-stamp. Hence, the user preference and fatigue vector only involve the behaviors so far. 

\subsubsection{Online Inference}

At this stage, advertisers launch service requests with (multiple sets of) campaign criteria, and AdVance returns the expected performances. We begin with building a simulated future campaign environment. Following the same methods as \cite{yahoo-cui2013campaign, facebook-jiang2015predicting, alibaba-chen2022unified}, we predict the number of impressions that satisfy the specified user targeting. We then sample auction records from the previous day's log according to the predicted volume. This offers more realistic competitor features and timely user interest. To better approximate the future environment, the Tencent platform delicately categorizes user targeting into 188,785 classes and utilizes CLOCK \cite{wang2023clock}, a multi-variable neural forecaster, to accurately predict the impression volume of each class. 

Once the future environment is built, we insert the target ad and its bid price into each auction. We feed the modified auction sequence to AdVance to re-calculate each auction's win-rate, expected yield, and the final campaign performance. To simulate interest evolution, we randomly append new ads to the user's display history according to win-rates and update the click sequence according to pCTR and pCVR. The fatigue vector is recurrently updated accordingly. After traversing the auction sequence with linear complexity, AdVance outputs the expected cost, exposure, click, and conversion volumes.

\section{Experiments}
\label{sec:section5}

We conduct experiments and ablation studies on industrial datasets from Tencent Advertising to validate our AdVance framework and investigate four research questions, \ie, \textbf{RQ1:} Prove AdVance's efficacy and superiority over state-of-the-art methods for campaign performance forecasting. \textbf{RQ2: }Demonstrate the necessity of modeling user preference and fatigue evolution. \textbf{RQ3: }Highlight the importance of introducing auxiliary tasks for auction representation and campaign-level prediction. \textbf{RQ4: }Evaluate the impact of different sequence-modeling techniques for campaign-level summarization. Lastly, we introduce the online A/B testing of AdVance to present its practical value in real-world scenarios. 

\subsection{Experimental Settings}

\subsubsection{Dataset}
We aim to train models that can integrate auction- and campaign-level information. Hence, the dataset should contain user history and each auction's competitor ads. No public dataset satisfies the requirements, so we collected our dataset from the Tencent Advertising platform. This dataset comprises 1.5 billion records from June 1, 2023, to June 30, 2023. Each record contains the user feature, user display history with corresponding behaviors, contextual information, and all competitor ads with their ad content, category ID, targeting criteria, bid price, etc. 

We focus on campaigns with over 20,000 records for better data quality and lower variance. Two business concerns also support this: First, advertisers with higher investments are more sensitive to budget efficiency. They are also more likely to increase investment when receiving positive feedback from AdVance. Second, ads with more frequent exposure in a longer period are often more prone to interest evolution and fatigue. We select 6,000 campaigns, with 1,000 for CPM, 2,000 for CPC, and 3,000 for CPA ads. 

\subsubsection{Compared Methods}
We compare with auction-level methods, which beat coarse-grained ones by a large margin. The baseline methods include those from the industry like Yahoo \cite{yahoo-cui2013campaign}, Microsoft \cite{microsoft-nath2013ad}, and Alibaba \cite{alibaba-chen2022unified}, and academic works as follows: 1) \textbf{CPF} \cite{yahoo-cui2013campaign} assumes a mixed log-normal distribution for bid prices and estimates its mean and standard deviation by regression. The win-rate is calculated by the cumulative density function (CDF). CPF trains decision trees to predict click/conversion rates and multiply them with the win-rates, thereby obtaining the expected yield. The final result is the accumulation of auction-level performance. 2) \textbf{GMIF} \cite{microsoft-nath2013ad} uses a first-order Dynamic Linear Model to forecast the number of future impressions. It then trains a Bayes net to estimate the threshold bid price to win a specific auction. Its paper omits the model design of pCTR/pCVR, so we use DeepFM \cite{guo2017deepfm} instead. 3) \textbf{MTLN} \cite{alibaba-chen2022unified} assumes static user traffic and uses DeepFM to estimate yields. Like us, MTLN also introduces a campaign-level model conditioned on the accumulated performance. An MMoE \cite{MMoE-ma2018modeling} model generates the final result. 4) \textbf{DLF} \cite{ren2019deep} surpasses previous works \cite{wu2018deep, lee2018deephit, MM-wu2015predicting, KM-zhang2016bid} in win-rate estimation. It discards the prior assumptions of win-rate distribution and uses a dedicated RNN to learn a discrete probability over the bid price. 5) \textbf{MTAE} \cite{yang2021multi} further enhances DLF with multi-task learning, leveraging the correlation between win-rates and click-through rates prediction. 

\subsubsection{Evaluation Metric}

The Tencent platform keeps storing new auction records and uses them to update numerous online models. The records form an ever-increasing time series, and we adopt a \textit{\textbf{sliding-window}} paradigm: trained on an input window of records, the model predicts the campaign performance for a future period (\aka, \textit{forecasting horizon}) of records. The window then goes on with a stepsize of 1 hour, and we fine-tune the model using new samples. This process is executed recurrently. At each time step, we calculate the weighted mean absolute percentage error (WMAPE):  
\begin{align}
    \textrm{WMAPE} \coloneqq \sum_i \textrm{weight}_i \cdot \frac{|y_i - \hat{y}_i|}{|y_i|},    
\end{align}
where $\textrm{weight}_i$ is the ratio between the $i$-th campaign's cost and the total cost of all campaigns, $y_i$ and $\hat{y}_i$ represent the ground truth and estimation, respectively. As a warm-up, we pre-train all models on the records from June 1, 2023, to June 7, 2023. Then, we accumulate the WMAPE per step and calculate the \textit{average} as the evaluation metric. We retrain the model on the whole dataset every 24 hours. We vary the forecasting horizon to evaluate the capacity of modeling long sequences as 1, 6, 12, and 24 hours. 

\subsubsection{Implementation Details}

We set the display history length to 300. Displayed items, fatigue vectors, user features, contextual information, and candidate ads are all 256-dimensional embeddings. We stack three encoder layers with four heads and 1024 hidden dimensions. $\theta_{win}$, $\theta_{CTR}$, and $\theta_{\overline{CVR}}$ are all three-layer MLPs with hidden neurons [128, 64, 1] and ReLU activation. We stack three SSM layers with the hidden state dimension $N=16$ for local and global modeling. The final campaign performance model $\theta_{CP}$ is an MLP of [128, 64, 4] with ReLU activation. The model is trained with an AdamW optimizer with a learning rate of 0.001, $\beta_1$ of 0.9, $\beta_2$ of 0.995, and $\epsilon$ of 1e-07. Batch-size = 32. Due to their equal value, the win-rate and yield prediction loss weights are set as [0.5, 0.5].  


\subsection{System Performance}

As shown in Table \ref{tab:WMAPE-1}, all models exhibit performance declines when the forecasting horizons are prolonged. This is mainly caused by the distribution shift of the campaign environment, such as newly introduced campaigns and old ones adjusting their criteria. Despite these declines, AdVance consistently outperforms the other methods by integrating auction- and campaign-level information and capturing interest evolution, addressing \textbf{RQ1}. 

\begin{table}[]
\caption{Timestep-averaged WMAPE of exposure, click, conversion, and cost for five baselines and AdVance \wrt different forecasting horizons from 1H to 24H. The best results are highlighted in bold.  }
\label{tab:WMAPE-1}
\begin{tabular}{cccccccc}
\hline
\multicolumn{2}{c}{Method}                                            & CPF   & GMIF  & MTLN  & DLF   & MTAE   & AdVance        \\ \hline
\multicolumn{1}{c|}{\multirow{4}{*}{1H}}  & \multicolumn{1}{c|}{exp} & 0.138 & 0.126 & 0.113 & 0.105 & 0.092  & \textbf{0.045}\\
\multicolumn{1}{c|}{}                     & \multicolumn{1}{c|}{clk}  & 0.159 & 0.153 & 0.158 & 0.131 & 0.125  & \textbf{0.061}\\
\multicolumn{1}{c|}{}                     & \multicolumn{1}{c|}{cvr}  & 0.171 & 0.173 & 0.164 & 0.158 & 0.135  & \textbf{0.099}\\
\multicolumn{1}{c|}{}                     & \multicolumn{1}{c|}{cost} & 0.154 & 0.149 & 0.147 & 0.129 & 0.119  & \textbf{0.075}\\ \hline
\multicolumn{1}{c|}{\multirow{4}{*}{6H}}  & \multicolumn{1}{c|}{exp} & 0.174 & 0.159 & 0.155 & 0.142 & 0.127  & \textbf{0.069}\\
\multicolumn{1}{c|}{}                     & \multicolumn{1}{c|}{clk}  & 0.201 & 0.217 & 0.193 & 0.134 & 0.130  & \textbf{0.090}\\
\multicolumn{1}{c|}{}                     & \multicolumn{1}{c|}{cvr}  & 0.275 & 0.283 & 0.266 & 0.247 & 0.215  & \textbf{0.149}\\
\multicolumn{1}{c|}{}                     & \multicolumn{1}{c|}{cost} & 0.228 & 0.212 & 0.219 & 0.196 & 0.147  & \textbf{0.116}\\ \hline
\multicolumn{1}{c|}{\multirow{4}{*}{12H}} & \multicolumn{1}{c|}{exp} & 0.193 & 0.183 & 0.205 & 0.157 & 0.132 & \textbf{0.092}\\
\multicolumn{1}{c|}{}                     & \multicolumn{1}{c|}{clk}  & 0.215 & 0.248 & 0.230 & 0.159 & 0.141  & \textbf{0.101}\\
\multicolumn{1}{c|}{}                     & \multicolumn{1}{c|}{cvr}  & 0.298 & 0.311 & 0.347 & 0.256 & 0.249  & \textbf{0.183}\\
\multicolumn{1}{c|}{}                     & \multicolumn{1}{c|}{cost} & 0.267 & 0.271 & 0.295 & 0.213 & 0.167  & \textbf{0.132}\\ \hline
\multicolumn{1}{c|}{\multirow{4}{*}{24H}} & \multicolumn{1}{c|}{exp} & 0.254 & 0.231 & 0.269 & 0.176 & 0.159  & \textbf{0.112}\\
\multicolumn{1}{c|}{}                     & \multicolumn{1}{c|}{clk}  & 0.273 & 0.268 & 0.258 & 0.162 & 0.187& \textbf{0.104}\\
\multicolumn{1}{c|}{}                     & \multicolumn{1}{c|}{cvr}  & 0.317 & 0.321 & 0.366 & 0.267 & 0.254  & \textbf{0.196}\\
\multicolumn{1}{c|}{}                     & \multicolumn{1}{c|}{cost} & 0.283 & 0.270 & 0.304 & 0.196 & 0.192& \textbf{0.145}\\ \hline
\end{tabular}
\end{table}

Among the compared methods, CPF's log-normal assumption of bid prices severely limits its capacity to model the complex competition among bidders. In contrast, Bayes net captures the probabilistic connections among factors contributing to the auction victory, leading to improved GMIF performance. MTLN's multi-task structure also considers the auction- and campaign-level information. However, MTLN discards the auction representation. Its global model can not handle the long sequence of auction records; it only takes the accumulated performance and coarse-grained campaign statistics. DLF surpasses the aforementioned methods by a large margin. It replaces the pre-defined win-rate distribution with a more flexible RNN, allowing DLF to adapt to the varying competition environment. Finally, MTAE outperforms DLF regarding reduced model complexity and maintenance overheads. MTAE improves accuracy by leveraging correlations between multiple tasks. 

In summary, a more flexible form of win-rate modeling and multi-task learning can promote forecasting accuracy significantly. However, the lack of modeling user interest evolution and campaign-level sequence leads to inferior performance.

\subsection{Ablation Study}

We conduct ablation studies involving five AdVance variants to assess each component's individual contributions and effectiveness: 1) \textbf{Static: }We discard the display history and assume a static user interest. 2) \textbf{Pref: }We preserve the clicked items for user preference and disregard the user fatigue. 3) \textbf{Aux: }We remove the auxiliary tasks of win-rates and pCTR and directly learn the auction representation. 4) \textbf{Accu: }We do not accumulate the auction-level results, making the campaign-level forecasting independent. 5) \textbf{Reg: }We assume auction and display sequences have regular time intervals. 

\begin{table}[]
\caption{Timestep-averaged WMAPE of exposure, click, conversion, and cost for five variants of AdVance \wrt different forecasting horizons from 1H to 24H. The best results are highlighted in bold.  }
\label{tab:WMAPE-2}
\begin{tabular}{cccccccc}
\hline
\multicolumn{2}{c}{Method}                                            & Static & Pref & Aux   & Accu & Reg & AdVance        \\ \hline
\multicolumn{1}{c|}{\multirow{4}{*}{1H}}  & \multicolumn{1}{c|}{exp} & 0.142& 0.124& 0.168& 0.117& 0.051& \textbf{0.045}\\
\multicolumn{1}{c|}{}                     & \multicolumn{1}{c|}{clk}  & 0.188& 0.153& 0.179& 0.146& 0.072& \textbf{0.061}\\
\multicolumn{1}{c|}{}                     & \multicolumn{1}{c|}{cvr}  & 0.201& 0.168& 0.205& 0.155& 0.115& \textbf{0.099}\\
\multicolumn{1}{c|}{}                     & \multicolumn{1}{c|}{cost} & 0.177& 0.149& 0.188& 0.135& 0.087& \textbf{0.075}\\ \hline
\multicolumn{1}{c|}{\multirow{4}{*}{6H}}  & \multicolumn{1}{c|}{exp} & 0.199& 0.175& 0.254& 0.162& 0.074& \textbf{0.069}\\
\multicolumn{1}{c|}{}                     & \multicolumn{1}{c|}{clk}  & 0.218& 0.181& 0.276& 0.180& 0.102& \textbf{0.090}\\
\multicolumn{1}{c|}{}                     & \multicolumn{1}{c|}{cvr}  & 0.276& 0.245& 0.319& 0.225& 0.158& \textbf{0.149}\\
\multicolumn{1}{c|}{}                     & \multicolumn{1}{c|}{cost} & 0.236& 0.199& 0.287& 0.213& 0.123& \textbf{0.116}\\ \hline
\multicolumn{1}{c|}{\multirow{4}{*}{12H}} & \multicolumn{1}{c|}{exp} & 0.251& 0.187& 0.344& 0.191& 0.116& \textbf{0.092}\\
\multicolumn{1}{c|}{}                     & \multicolumn{1}{c|}{clk}  & 0.268& 0.206& 0.372& 0.218& 0.128& \textbf{0.101}\\
\multicolumn{1}{c|}{}                     & \multicolumn{1}{c|}{cvr}  & 0.379& 0.261& 0.401& 0.279& 0.192& \textbf{0.183}\\
\multicolumn{1}{c|}{}                     & \multicolumn{1}{c|}{cost} & 0.327& 0.223& 0.392& 0.245& 0.143& \textbf{0.132}\\ \hline
\multicolumn{1}{c|}{\multirow{4}{*}{24H}} & \multicolumn{1}{c|}{exp} & 0.325& 0.249& 0.466& 0.272& 0.117& \textbf{0.112}\\
\multicolumn{1}{c|}{}                     & \multicolumn{1}{c|}{clk}  & 0.317& 0.273& 0.501& 0.319& 0.123& \textbf{0.104}\\
\multicolumn{1}{c|}{}                     & \multicolumn{1}{c|}{cvr}  & 0.405& 0.312& 0.529& 0.355& 0.215& \textbf{0.196}\\
\multicolumn{1}{c|}{}                     & \multicolumn{1}{c|}{cost} & 0.382& 0.280& 0.485& 0.318& 0.171& \textbf{0.145}\\ \hline
\end{tabular}
\end{table}

As depicted in Table \ref{tab:WMAPE-2}, \textbf{Static}'s performance drops severely and further declines with a longer horizon. \textbf{Pref} alleviates such degradation by considering preference evolution, but its incomplete view of user interest makes it inferior to AdVance, thus addressing \textbf{RQ2}. The absence of supervision signals during model training imposes significant difficulty in learning a meaningful representation. This accounts for \textbf{Aux}'s performance gap. Using the \textit{Divide \& Conquer} policy, we decompose the campaign performance into numerous auction performances and accumulate them. The accumulated results serve as an informative reference and reduce the overall difficulty of forecasting. This improves \textbf{Accu}'s accuracy and addresses \textbf{RQ3}. The changes in traffic affect the supply of user impressions and stimulate the intensity of auctions. \textbf{Reg} omits the non-stationary traffic, causing an accuracy drop. 

\begin{table}[]
\caption{Timestep-averaged WMAPE results. Run on a V100-16GB GPU. OOM indicates \textit{out-of-memory}.  }
\label{tab:WMAPE-3}
\begin{tabular}{ccccccc}
\hline
\multicolumn{2}{c}{Method}                                            & Ind   & RNN   & Transformer          & S4    & AdVance        \\ \hline
\multicolumn{1}{c|}{\multirow{4}{*}{1H}}  & \multicolumn{1}{c|}{exp}  & 0.062& 0.061& \textbf{0.043}       & 0.059 & 0.045\\
\multicolumn{1}{c|}{}                     & \multicolumn{1}{c|}{clk}  & 0.071& 0.075& \textbf{0.061}       & 0.072 & \textbf{0.061}\\
\multicolumn{1}{c|}{}                     & \multicolumn{1}{c|}{cvr}  & 0.105& 0.102& 0.103                & 0.111 & \textbf{0.099}\\
\multicolumn{1}{c|}{}                     & \multicolumn{1}{c|}{cost} & 0.081& 0.079& 0.076& 0.081 & \textbf{0.075}\\ \hline
\multicolumn{1}{c|}{\multirow{4}{*}{6H}}  & \multicolumn{1}{c|}{exp}  & 0.094& 0.104& \textbf{0.066}& 0.083 & 0.069\\
\multicolumn{1}{c|}{}                     & \multicolumn{1}{c|}{clk}  & 0.113& 0.115&                      0.091& 0.104 & \textbf{0.090}\\
\multicolumn{1}{c|}{}                     & \multicolumn{1}{c|}{cvr}  & 0.158& 0.177&                      \textbf{0.147}& 0.162 & 0.149\\
\multicolumn{1}{c|}{}                     & \multicolumn{1}{c|}{cost} & 0.124& 0.146&                      \textbf{0.115}& 0.131 & 0.116\\ \hline
\multicolumn{1}{c|}{\multirow{4}{*}{12H}} & \multicolumn{1}{c|}{exp}  & 0.115& 0.133& 0.093& 0.103 & \textbf{0.092}\\
\multicolumn{1}{c|}{}                     & \multicolumn{1}{c|}{clk}  & 0.132& 0.145&                      0.107& 0.117 & \textbf{0.101}\\
\multicolumn{1}{c|}{}                     & \multicolumn{1}{c|}{cvr}  & 0.207& 0.242&                      0.184& 0.189& \textbf{0.183}\\
\multicolumn{1}{c|}{}                     & \multicolumn{1}{c|}{cost} & 0.147& 0.209&                      0.135& 0.148& \textbf{0.132}\\ \hline
\multicolumn{1}{c|}{\multirow{4}{*}{24H}} & \multicolumn{1}{c|}{exp}  & 0.150& 0.172& \multirow{4}{*}{OOM} & 0.128 & \textbf{0.112}\\
\multicolumn{1}{c|}{}                     & \multicolumn{1}{c|}{clk}  & 0.166& 0.181&                      & 0.149 & \textbf{0.104}\\
\multicolumn{1}{c|}{}                     & \multicolumn{1}{c|}{cvr}  & 0.205& 0.215&                      & 0.201 & \textbf{0.196}\\
\multicolumn{1}{c|}{}                     & \multicolumn{1}{c|}{cost} & 0.181& 0.186&                      & 0.172& \textbf{0.145}\\ \hline
\end{tabular}
\end{table}

To answer \textbf{RQ4}, we modify AdVance's global summarizer and obtain four variants: 1) \textbf{Ind: }No global model, using accumulated auction performance. 2) \textbf{RNN: }Using LSTM to summarize auction sequence. 3) \textbf{Transformer: }Using a quadratic-complexity encoder with time-stamped position embedding. 4) \textbf{S4: }Using an SSM with parameters independent of inputs and time intervals. 

As depicted in Table \ref{tab:WMAPE-3}, the lack of a holistic view of the campaign environment leads to \textbf{Ind}'s performance drop, proving the necessity of a global model. \textbf{RNN} features linear complexity but has difficulty memorizing too long context. In contrast, \textbf{Transformer} explicitly preserves all previous tokens, achieving the best performance for a moderate context length. However, the inference memory grows linearly \wrt input length and reports OOM for the 24-hour horizon. It also presents more than five times the latency of AdVance due to its quadratic complexity, making it unsuitable for online service.  \textbf{S4} \cite{gu2022S4} can compress long sequences with linear-complexity calculation, but it cannot selectively store salient information from an unevenly distributed time series, leading to its accuracy declines.  

In summary, a comprehensive solution for CPF tasks should consider the user's preference \& fatigue evolution, long-context modeling with low complexity, and a tight connection between auction- and campaign-level information. 

\subsection{Further Investigation}
AdVance's user interest and local auction modules constitute a click-through rate model. We compare it with representative pCTR models to demonstrate the impact of interest evolution, especially for long-period campaigns: 1) \textbf{Wide\&Deep} \cite{widedeep-cheng2016wide}: A combination of a deep neural network and a linear model that captures low- and high-order feature interactions. 2) \textbf{DIEN} \cite{dien-zhou2019deep}: A sequential model that considers interactions between user clicks and candidate ads. A dedicated RNN captures the evolution of user preference over time. 3) \textbf{FAN} \cite{FAN-li2023fan}: An improved interest model that incorporates the frequency-domain feature for user fatigue. 

\begin{figure}
    \centering
    \includegraphics[width=0.85\linewidth]{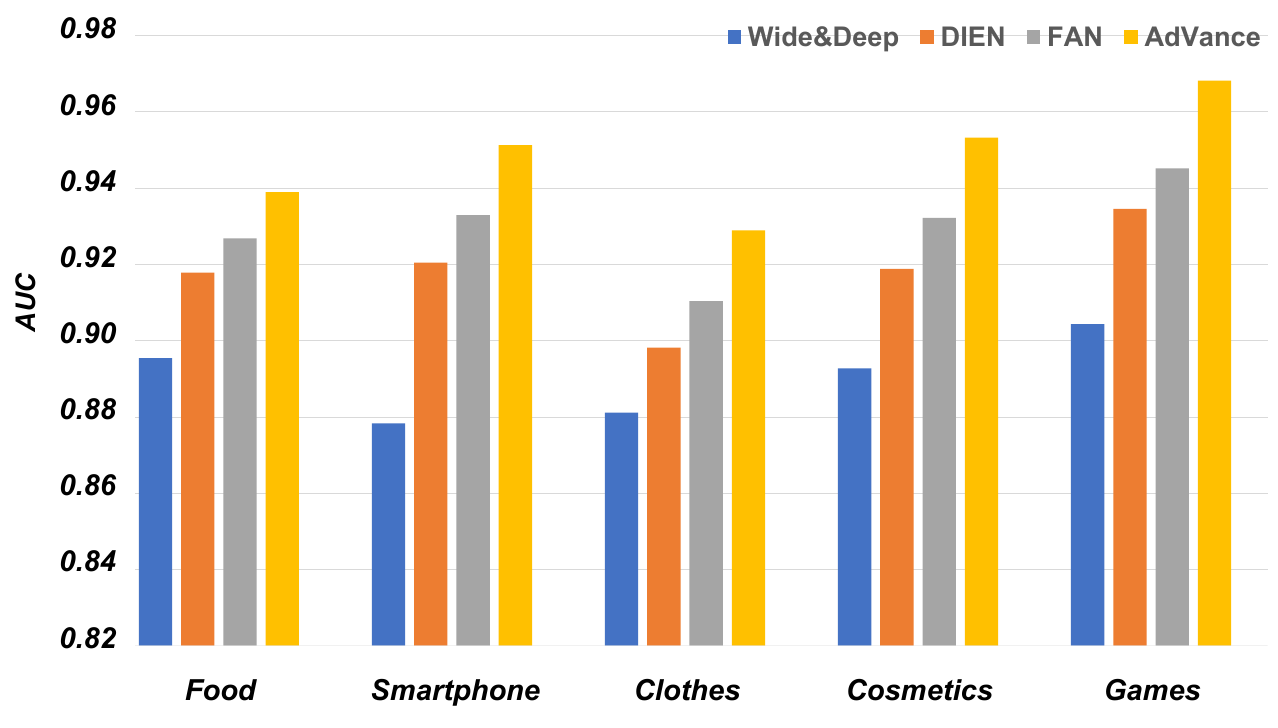}
    \caption{The AUC of three baselines and AdVance on five campaigns from various industries. }
    \label{fig:auc}
\end{figure}

We select five campaigns from various industries, \ie, food, smartphones, clothes, cosmetics, and games. For each campaign, we use 80\% records as the training set and 20\% as the testing set. The results are shown in Fig. \ref{fig:auc}. \textbf{Wide\&Deep} performed the worst as it only considers static user features and can not model the user's sequential behaviors. In contrast, \textbf{DIEN} performed better by capturing user preferences hidden in clicked items. \textbf{FAN} calculates the fast Fourier transformation of the displayed yet non-clicked items to model user fatigue. However, FAN assumes regular time intervals, and FFT features are inadequate to represent interest evolution compared to deep neural networks. 

In conclusion, both clicked and non-clicked items are necessary to capture the evolution of preference and fatigue, significantly affecting yield prediction accuracy.

\subsection{Online A/B Testing}

Our AdVance has been implemented on the Tencent Advertising platform, allowing advertisers to try various campaign criteria and receive corresponding performances in real-time. Advertisers can then decide the appropriate campaign settings based on predictions and business demands. Given specific criteria, we use the inverted index to retrieve qualified records from the system log of the past 24 hours with a maximum number of 20,000 records, ensuring a response delay within 5 seconds. We scale the prediction accordingly to maintain consistency. 

To evaluate AdVance's effectiveness, we conducted online A/B tests on advertisers of the same industry. We split them into two groups with similar \textit{average revenue per user} (ARPU). Only advertisers in group B were granted access to the AdVance services. Over two weeks, the comparison revealed a 4.5\% uplift in ARPU for group B advertisers due to optimized campaign configuration. AdVance is processing thousands of queries daily, greatly enhancing the platform's income and attractiveness to advertisers. 

\section{Discussion and Future Work}

Like many other industrial practices, AdVance mainly considers user traffic fluctuation when modeling environment dynamics and handles it with a fine-grained time series model. However, new campaigns may participate, and other advertisers may adjust their bid prices or user targeting as a counterbalance. This can cause a drop in accuracy over long periods, shown in Table \ref{tab:WMAPE-1}. 

One possible mitigation is introducing advertiser modeling techniques 
and game-theory-based competition modeling \cite{10.1145/2488388.2488513, 10.1145/3488560.3498479}. The former can help predict when and how new campaigns will be launched, and the latter can predict the possible response from competitors. These future directions hold promise for advancing the field of ad campaign performance forecasting and facilitating more effective decisions in online advertising.
\section{Conclusion}

We propose AdVance, a time-aware framework integrating auction- and campaign-level modeling. We introduce user preference as a time-positioned click sequence and emphasize fatigue modeling by compressing all displayed history into a concise vector. We trained an encoder in a supervised manner to predict cost and yield per auction. The encoder applies self-attention/cross-attention on candidate ads and user features, thereby converting each auction into informative embedding. To comprehend the generated long, irregular sequence, we make the linear-complexity SSM's parameters dependent on current embedding and time interval. The conditional SSM then outputs expected campaign performance, with its prediction conditioned on the accumulation of auction-level results. AdVance outperforms state-of-the-art methods on large-scale industrial datasets, and has been deployed on the Tencent advertising system, showing a 4.5\% uplift in Average Revenue per User in the A/B test.

\newpage
\balance
\bibliographystyle{ACM-Reference-Format}
\bibliography{Reference}

\clearpage
\begin{appendices}

\setcounter{table}{0} 
\setcounter{figure}{0} 
\setcounter{equation}{0} 

\renewcommand{\thetable}{\thesection-\arabic{table}} 
\renewcommand{\theequation}{\thesection-\arabic{equation}} 
\renewcommand{\thefigure}{\thesection-\arabic{figure}}
\section{Tencent Advertising Platform}
\nobalance

This section offers more background knowledge of the Tencent advertising platform on which AdVance has been implemented, including auction workflow, data log, and filtering rules. 

\subsection{Real-time Bidding Workflow}

\begin{figure}[htb]
    \centering
    \includegraphics[width=0.95\linewidth]{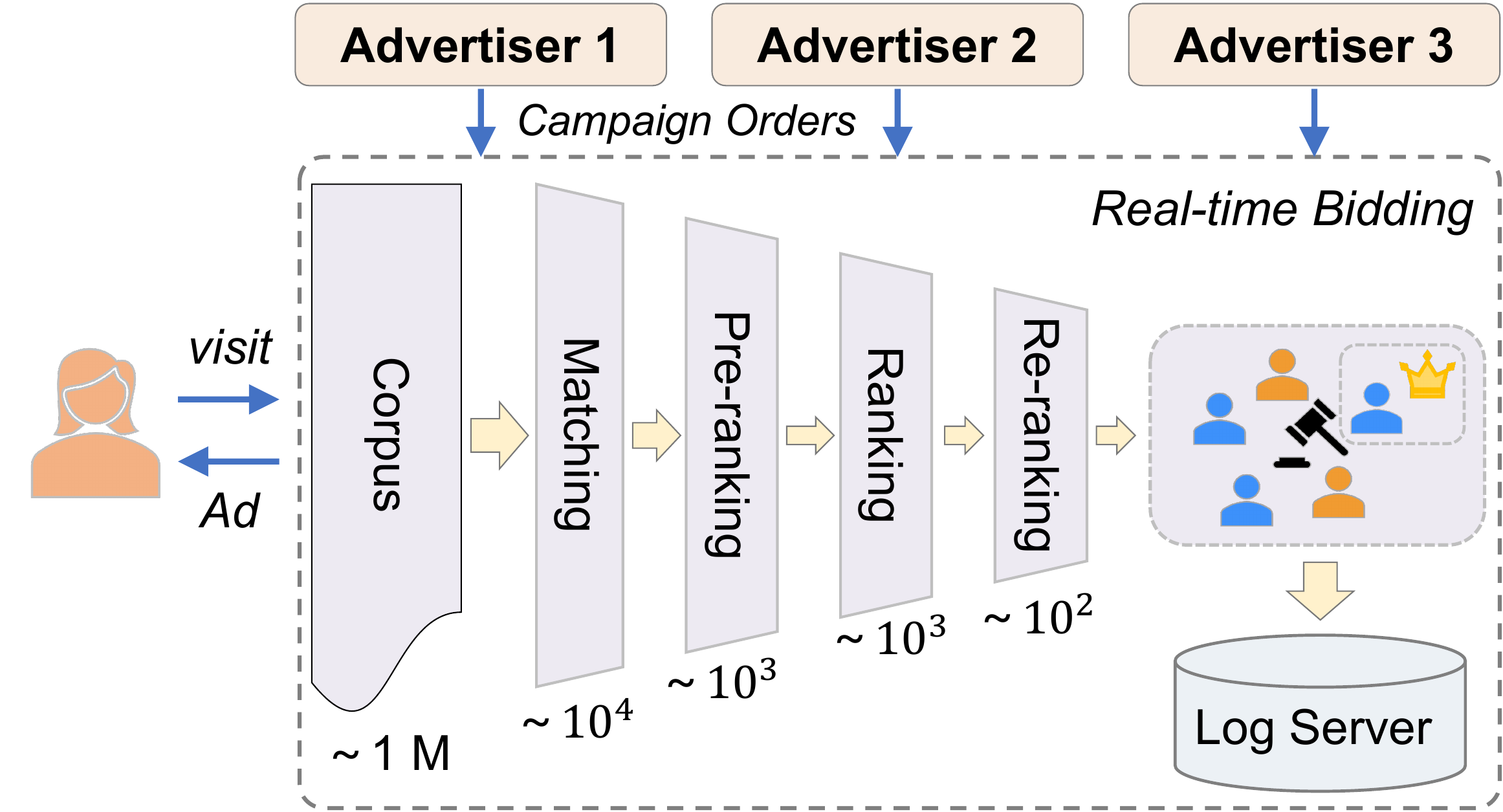}
    \caption{Real-time bidding workflow and the funnel-shaped structure. }
    \label{fig:rtb_procedure}
\end{figure}

Whenever a user visits Tencent's platforms (\eg, Tencent Video or Tencent News), an opportunity of showing an advertisement emerges. We name such opportunities as \textit{impressions} and sell them to advertisers for revenue. For each impression, the ad platform retrieves relevant ads with matched campaign criteria and initiates an auction. As shown in Fig. \ref{fig:rtb_procedure}, the platform adopts a funnel-shaped structure to handle millions of ads in the corpus, including matching, pre-ranking, ranking, and re-ranking phases. This structure strikes a balance between precise ad retrieval and timely response. Each phase progressively reduces the number of candidate ads and employs more complex and accurate algorithms. Finally, about 200 candidates can participate in the re-ranking competition, and the decision-making process considers user-ad matching, bid price, and the platform's own strategy.

\subsection{Data Log and Embedding}
\label{embedding}
The Tencent platform stores auction records in a log server to support various data-driven algorithms such as pCTR/pCVR estimation and campaign modeling. Each record contains multi-source information: 

\begin{itemize}
    \item \textbf{User features}: age, gender, location, device type,  \etc
    \item \textbf{Contextual information}: ad slot placement (web, app), content topic category, timestamp,  \etc
    \item \textbf{History}: ad content, ad category ID, corresponding user behaviors (click or purchase),  \etc
    \item \textbf{Candidate ads}: ad creative ID, campaign criteria (user attributes, demographics, keywords), ad type (CPM/CPC/CPA), bid price, auction winner, \etc
\end{itemize}

Note that the platform adopts a down-sampling strategy to handle the overly large user queries (often billion-level per second), \ie, only the auctions of a particular group of user IDs are recorded.  User IDs are obtained by uniform sampling from the total ID dictionary. The ratio depends on the I/O and computation capacity of the log servers. 

The recorded features can be categorized into continuous (\eg, age, timestamp) and categorical (\eg, gender, location) features. We convert the auction records into a set of fixed-length embedding. Specifically, each categorical feature is represented as a vector of one-hot encoding, and each continuous feature is represented as the value itself. One-hot vectors are extremely sparse, so we employ a domain-specific embedding layer to compress them to a low-dimensional, dense vector before feeding into the model. Such an embedding layer is dedicated to each feature domain to lower the total parameter volume. Finally, we concatenate these vectors to obtain the corresponding user feature, context, and candidate ads embeddings as model input shown in Fig. \ref{fig:AdVance}. 

\subsection{Manual Filtering Rules}

\begin{figure}[h]
    \centering
    \includegraphics[width=0.95\linewidth]{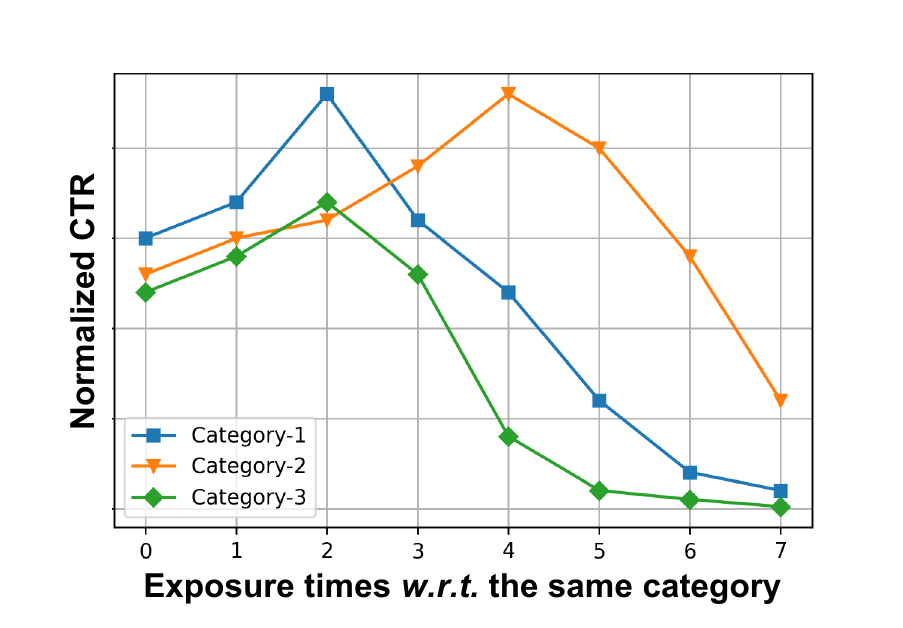}
    \caption{The CTR trends vary with the number of exposures to ads of the same category. We present three categories, all showing a decline when over-exposed. We normalize the data for business privacy. }
    \label{fig:ctr_trend}
\end{figure}

In Sec. \ref{user_interest}, we adopt a data-driven method to capture user interest evolution, which can be visualized in Fig. \ref{fig:ctr_trend}. However, the platform must consider various factors in the real-world business scenario to satisfy advertiser demands and enhance long-term user experience. These factors make the auction process more than a simple bid ranking problem. Hence, Tencent manually defines multiple filtering rules in the re-ranking phase to discard certain ads as a post-process, including but not limited to 

\begin{itemize}
    \item \textbf{Budget Pacing:} Ensures budget is spent evenly throughout the campaign period, avoiding front-loading or overspending. 

    \item \textbf{Frequency Capping:} Limits the number of times a user sees the same ad or ads from the same industry, preventing ad fatigue and maximizing reach. 

    \item \textbf{Brand Safety:} Protects advertisers from their ads appearing alongside inappropriate or harmful content.
\end{itemize}

These filtering rules are designed based on human experience and can not be described by analytic functions to insert into models. Therefore, we employ a supervised training paradigm with the data log to approximate the effect of such rules on the auction process. 

\section{Baseline Setting}

This section offers more details of the implementation of the baseline methods. 

\subsection{Setting of Compared Methods}

\textbf{CPF} \cite{yahoo-cui2013campaign}: We adopt a mixture of two log-normal distribution. The density function is $g(x;\mu_1, \sigma_1, \mu_2, \sigma_2, p)=(1-p)f(x;\mu_1, \sigma_1) + pf(x;\mu_2, \sigma_2)$ and we set $p$ as 0.1. We train a Factorization-Machine \cite{rendle2010FM} on the feature embeddings introduced in App. \ref{embedding} to estimate $\mu_1, \sigma_1, \mu_2, \sigma_2$. We adopt the classical XGBoost model \cite{chen2016xgboost} to predict the pCTR/pCVR. We use the accumulated yield and cost as the final campaign performance. 

\textbf{GMIF} \cite{microsoft-nath2013ad}: For the impression forecasting part, we train a DLM for each user attribute at the hour level. We follow the original recursive function in the paper but change its parameters to $W = 5, V = 15, C_0=20$. To estimate the threshold price of winning the auction, we train a Bayes net to estimate the conditional probability between input variables. Here, each variable corresponds to one feature domain, such as age, gender, location, \etc Its paper omits the model design of pCTR/pCVR, so we use DeepFM \cite{guo2017deepfm} trained on the feature embeddings introduced in App. \ref{embedding}. We use the accumulated yield and cost as the final campaign performance.

\textbf{MTLN} \cite{alibaba-chen2022unified}: We train a DeepFM to estimate the pCTR/pCVR for each auction and multiply the result with the bid price to obtain the eCPM. We compare the calculated eCPM with the threshold price of each auction record and accumulate the yield. We feed the accumulated yield and campaign-level statistics into an MMoE model consisting of four Expert-MLPs of [128, 64], four Tower-MLPs of [64, 32, 1] with ReLU activation, and one gate model of MLP [64, 4]. The four Tower-MLPs correspond to the cost, impression, click, and conversion volumes. 

\textbf{DLF} \cite{ren2019deep}: We discretize the scale of bid price into 100 sub-intervals. As the number of candidate ads varies, we feed our auction representation as the DLF's input. We stack two layers of LSTM with a hidden dimension of 512 to predict the conditional win-rate for each auction. The pCTR and pCVR are estimated using the same DeepFM model as MTLN. We use the accumulated yield and cost as the final campaign performance. 

\textbf{MTAE} \cite{yang2021multi}: We feed our auction representation as the MTAE's input. MTAE adopts a multi-task paradigm and two top-models share the input: one consists of MLPs as our $f(\cdot; \theta_{CTR})$ and $f(\cdot; \theta_{\overline{CVR}})$, and the other is an MLP of [256, 100] for estimating a discrete distribution over threshold price. Here, we again evenly divide the scale of the bid price into 100 sub-intervals. MTAE further superimposes the DLF model over the bid price model as an auxiliary task. We use the accumulated yield and cost as the final campaign performance. 

\subsection{Setting of Ablation Study}

\textbf{RNN}-variant stacks three layers of LSTM with a hidden dimension of 512. \textbf{Transformer}-variant stacks three encoder layers with four heads and a hidden dimension of 1024 (expansion = 4). \textbf{S4}-variant stacks three layers of SSM with $N=16$ using the authors' open-source code.  

\subsection{Setting of Further Investigation}

We follow the original structure of \textbf{Wide\&Deep} \cite{widedeep-cheng2016wide}, where the deep model is an MLP of [128, 64, 32] on our auction representation, and the wide model is a generalized linear model on the one-hot vector in App. \ref{embedding}. For the \textbf{DIEN} \cite{dien-zhou2019deep} model, we employ a two-layer Gated Recurrent Network (GRN) with a hidden dimension of 512 to capture the interest evolution. The obtained interest vector is concatenated with user features, context, and target ad embedding. Then we feed it into an MLP of [256, 128, 1] to predict the pCTR. For the \textbf{FAN} \cite{FAN-li2023fan} model, we set the length of \textit{N}-point FFT as 300 and keep the other settings unchanged as the original paper. 

\end{appendices}

\clearpage
\end{document}